# Spin-Triplet Excitonic Insulator in the Ultra-Quantum Limit of HfTe$_5$

Jinyu Liu[1], Varsha Subramanyan[2,3], Robert Welser[1], Timothy McSorley[1], Triet Ho[4], David Graf[5], Michael T. Pettes[3], Avadh Saxena[2], Laurel E. Winter[6], Shi-Zeng Lin[2,3], Luis A. Jauregui[1,*]

[1]Department of Physics and Astronomy, The University of California, Irvine, CA 92697, USA
[2]Theoretical Division T-4, Los Alamos National Laboratory, Los Alamos, NM 87545, USA
[3]Center for Integrated Nanotechnologies (CINT), Los Alamos National Laboratory, Los Alamos, NM 87545, USA
[4]Department of Mechanical and Aerospace Engineering, The University of California, Irvine, CA 92697, USA
[5]National High Magnetic Field Laboratory, Florida State University, Tallahassee, FL 32310, USA
[6]National High Magnetic Field Laboratory, Los Alamos National Laboratory, Los Alamos, New Mexico 87544, USA
* Corresponding email: lajaure1@uci.edu

## Abstract

More than fifty years ago, excitonic insulators, formed by the pairing of electrons and holes due to Coulomb interactions, were first predicted [1–3]. Since then, excitonic insulators have been observed in various classes of materials, including quantum Hall bilayers, graphite, transition metal chalcogenides, and more recently in moiré superlattices. In these excitonic insulators, an electron and a hole with the same spin bind together and the resulting exciton is a spin singlet. Here, we report the experimental observation of a spin-triplet exciton insulator in the ultra-quantum limit of a three-dimensional topological material HfTe$_5$. We observe that the spin-polarized zero[th] Landau bands, dispersing along the field direction, cross each other beyond a characteristic magnetic field in HfTe$_5$, forming the one-dimensional Weyl mode. Transport measurements reveal the emergence of a gap of about 250 μeV when the field surpasses a critical threshold. By performing the material-specific modeling, we identify this gap as a consequence of a spin-triplet exciton formation, where electrons and holes with opposite spin form bound states, and the translational symmetry is preserved. The system reaches charge neutrality following the gap opening, as evidenced by the zero Hall conductivity over a wide magnetic field range (10 - 72 T). Our finding of the spin-triplet excitonic insulator paves the way for studying novel spin transport including spin superfluidity, spin Josephson currents, and Coulomb drag of spin currents in analogy to the transport properties associated with the layer pseudospin in quantum Hall bilayers.

At high magnetic fields ($B$), the motion of charged particles within two-dimensional (2D) or three-dimensional (3D) systems becomes quantized into discrete energy levels known as Landau levels (LLs). As $B$ increases, the system reaches the quantum limit (QL), where charge carriers primarily occupy the lowest LL(s), or the zero[th] LL(s) in the case of relativistic fermions. In this regime, quenched kinetic energy and a significantly enhanced density of states (DOS) give rise to exotic phenomena. In 2D, the interplay between electron-electron interactions and the topological nature of LLs can lead to correlated electronic phases, including Wigner crystals [4], charge density waves [5,6], excitonic insulators [7,8], and the fractional quantum Hall effect [9,10]. In 3D, while the energy-momentum relation remains dispersive along the field direction, the electron motion perpendicular to *B* is confined, effectively reducing the dimensionality to one-dimensional (1D). This confinement, coupled with quenched kinetic energy, theoretically promotes interaction-driven instabilities similar to those in 2D systems [11,12]. However, due to the required high fields to reach the ultra-QL in conventional metals, identifying such phases in 3D systems remains challenging, with evidence limited to graphite [13–17] and bismuth [18].

Recent discoveries in topological materials offer new avenues to explore QL phenomena in 3D systems [19–24]. Transition metal pentatellurides, *i.e.* (Zr, Hf)Te$_5$, have attracted attention due to their nontrivial topological characteristics and unusual transport properties [25–28]. Their small Fermi surfaces make them highly suitable for exploring novel quantum phenomena in the ultra-QL. For instance, as *B* increases beyond the QL ($B_{QL}$), the evolution of the cyclotron energy and the Landau band (LB) splitting can lead to the zero$^{th}$ LB crossing, particularly significant in the weak topological insulator (WTI) phase [29,30]. With large Landé g-factors, the Zeeman energy can surpass both the bandgap and the Fermi energy, driving the spin-polarized zero$^{th}$ LBs of the conduction ($0^-$) and valence ($0^+$) bands toward each other, as depicted in the schematic of Fig. 1(a) (i). This leads to inter-LB crossings, persistently closing the bandgap [29–31], as shown in Fig. 1(a) (ii). The chemical potential (μ) is modulated by *B* and beyond a characteristic field *B\**, a 1D Weyl mode is created, effectively generating an anisotropic 1D band structure without geometric confinement (Fig. 1(a) (iii & iv)).

Furthermore, the electronic liquid in 1D systems is well known for its susceptibility to forming various correlated quantum states [32]. When *μ* approaches the 1D Weyl nodes, the enhanced DOS can further promote correlated phenomena. One potential phase that may emerge is the excitonic insulator (EI), as predicted by Kohn in small-gap semiconductors where the exciton binding energy exceeds the bandgap. This state forms through a coherent pairing of electrons and holes, exhibiting a charge-neutral Bardeen-Cooper-Schrieffer (BCS)-like condensation [3]. Particularly, spin-triplet EIs can exhibit intriguing magnetic properties, such as spin superfluidity, analogous to spin-triplet superconductivity [33,34] and the superfluidity of $^3$He [35]. However, achieving such spin-triplet EI remains a significant challenge [36–38]. Interestingly, the electron and hole zero$^{th}$ LBs in HfTe$_5$ resemble a direct-gap semiconductor with a tunable bandgap, modulated by *B*. These spin-polarized zero$^{th}$ LBs allow the creation of spin-triplet excitons, where the electron and hole bands have opposite spins, enabling a spin-triplet EI phase [36]. Therefore, HfTe$_5$ is an ideal material candidate to explore the magnetic field-driven spin-triplet EIs or other correlated states.

Here, we investigate quantum transport properties in the ultra-QL of low-carrier-density HfTe$_5$ samples, revealing three key phenomena. First, a Lifshitz transition to the 1D Weyl phase occurs as spin-polarized $0^+$ and $0^-$ LBs cross with each other [29–31], leading to a negative longitudinal magnetoresistance (NLMR) in the interlayer resistivity ($\rho_{zz}$). Second, a field-driven metal-to-insulator transition is observed, characterized by a gap opening (~ 250 μeV), as indicated by intralayer ($\rho_{xx}$) resistivity and $\rho_{zz}$ measurements. Third, the Hall resistivity ($\rho_{xy}$) vanishes demonstrating a charge-neutral state beyond a characteristic field ($B_{CNP}$), with $B_{CNP}$ exhibiting a BCS-like temperature dependence. This charge-neutral state persists to the highest measured fields. Material-specific modeling suggests that this insulating state arises from electron-hole pairing, forming a spin-triplet EI gap at the 1D Weyl nodes, supported by the zero Hall conductivity and nonlinear transport measurements. These findings highlight interaction-driven quantum phenomena in the ultra-QL of HfTe$_5$, advancing our understanding of correlation effects in topological materials.

Our chemical vapor transport (CVT) grown HfTe$_5$ samples are elongated, belt-like single crystals with an orthorhombic structure belonging to the *cmcm* space group. Previous studies [29,39] have identified pristine HfTe$_5$ as a WTI with a Dirac gap of ~3 meV at the Γ point and highly anisotropic band dispersions along different momentum directions. For the intralayer measurements, current (*I*) is applied along the *a*-axis with *B* perpendicular to the *ac*-plane. Figure 1(b) shows $\rho_{xx}$ and $\rho_{xy}$ vs. *B* at $T = 0.3$ K, for sample #1 (S#1). In the low field regime, $\rho_{xx}$ exhibits Shubnikov–de Haas (SdH) oscillations with a frequency of 1.2 T. Meanwhile, $\rho_{xy}$ varies linearly with *B*, corresponding to an electron density of ~ $10.2 \times 10^{16}$ cm$^{-3}$. Similar to the 3D quantum Hall effect observed in (Zr, Hf)Te$_5$ [27,40], $\rho_{xy}$ in our samples displays plateaus with corresponding valleys in $\rho_{xx}$, allowing the identification of LLs. Beyond $B_{QL} = 1.3$ T, the system enters the QL, where the 1$^{st}$ LL is depopulated, and only the zero$^{th}$ LLs are occupied.

Figure 1(c) shows the calculated evolution of the zero$^{th}$ LLs, μ, and the charge neutrality point (CNP) at $k_z = 0$ as function of *B*. Our simulations suggest that μ crosses the 0$^+$ LL at ~10 T and the CNP at $B_{CNP}$ ~ 25 T, using the carrier density from S#1 in good agreement with the experimental result (Fig. S4). Figure 1(d) shows $\sigma_{xx}$ and $\rho_{xy}$ vs. *B* revealing a broad peak in $\rho_{xy}$ far beyond $B_{QL}$, coinciding with a peak in $\sigma_{xx}$ which is proportional to the DOS. The field at this peak, $B^*$ ~ 10 T, marks the onset of a hole introduction. For $B > B^*$, as shown in Fig. 1(a), μ crosses both 0$^-$ and 0$^+$ LBs, and the transport is dominated by the 1D Weyl mode.

To corroborate the existence of 1D Weyl mode transport, we perform interlayer current measurements on another sample, S#2, by placing multiple electrodes on the top and bottom surfaces, as shown Fig. 2(a). A uniform current density across the *ac*-plane is ensured as simulated by the finite element analysis performed with COMSOL Multiphysics (Fig. 2(b)). Figure 2(c) illustrates the formation of the 1D Weyl state in real space for $B > B^*$. With our measurement setup, the parallel electric (*E*) and *B* fields condition is fulfilled by applying *I* and *B* both along the sample's *b*-axis, producing imbalanced chirality of 1D Weyl fermions. Figure 2(d) shows $\rho_{zz}$ vs. *B* at different temperatures under pulsed fields. At $T = 0.6$ K, $\rho_{zz}$ rises with *B*, reaching a peak at $B \sim 9.5$ T. Beyond this field, $\rho_{zz}$ decreases, reaching a minimum around 35 T. The *B* onset of the NLMR agrees well with the calculated and measured $B^*$ ~ 10 T from S#1 & S#2. Moreover, $\rho_{zz}$ exhibits a 1/$B^2$ dependence, as shown by the fitting lines in Fig. 2(d). The NLMR is sensitive to the orientation of *B* and vanishes when *B* is tilted beyond 44° (Fig. S3). Therefore, the observed NLMR and the *E·B*-like characteristic, similar to the 3D counterpart, unambiguously demonstrates the existence of the 1D Weyl mode. The NLMR persists up to $T = 10$ K and vanishes at $T = 20$ K, likely due to the occupation of higher LLs at elevated temperatures.

We note that for $B > 40$ T, $\rho_{zz}$ starts to increase with *B* as shown in Fig. S3(a), indicating the presence of a competing mechanism. Additionally, the temperature dependence of $\rho_{zz}$ reveals the sample is insulating for $T < 10$ K at all measured fields. One possible explanation for this behavior is discussed below with the introduction of a correlated gap, leading to the disappearance of the NLMR, similarly observed in known type-I Weyl semimetals such as TaAs [21].

To better characterize the QL transport behaviors, we measured sample S#3, characterized by a smaller electron density ($\sim 5.18 \times 10^{16}$ cm$^{-3}$; see Fig. S2). The SdH oscillations with a frequency of 1.0 T results in $B_{QL} = 0.9$ T. The onset of 1D Weyl mode transport, marked by the extremum in $\rho_{xy}$, occurs at $B^* = 6$ T. Furthermore, a metal-to-insulator transition is observed at a critical field $B_C \sim 6$ T, as indicated by $\rho_{xx}$ vs. $B$ measured at different temperatures (Fig. S4(c)). This metal-to-insulator transition is consistently observed across multiple samples (Figs. S4 and S8(a)), with $B_c$ decreasing for samples with lower electron densities (Fig. S9(a)).

S#3 was further investigated under pulsed fields (Fig. 3). For $T < 4$ K, $\rho_{xx}$ rapidly increases with $B$ for $B < 20$ T, transitioning to a slower increase at higher fields (Fig. 3(a)). For $\rho_{xy}$, deviations from linear field dependence occur beyond $B_{QL}$, as shown in Fig. 3(b). Remarkably, $\rho_{xy}$ reaches a maximum near $B^* = 6$ T, before decreasing rapidly and crossing zero at the CNP around $B_{CNP} \sim 11$ T. For $B > B_{QL}$ and $T < 4$ K, $\rho_{xy}$ remains near zero, indicating a transition from electron-dominated behavior at low fields to a state of equal electron-hole density up to $B = 72$ T. However, for $T = 4$ & 10 K, $\rho_{xy}$ does not reach zero, suggesting that the carrier type remains electron-dominated at all fields, preventing μ from reaching the CNP. For $B > B_C$, the temperature dependence of $\rho_{xx}$ is well described by the Arrhenius equation, $\rho_{xx} \propto \exp(\frac{\Delta}{2k_BT})$, where the thermal activation energy ($\Delta$) increases with $B$. $\Delta$ reaches $\sim 250$ μeV at $B \sim 20$ T, and stabilizes around 225 μeV at higher fields (Fig. 3(c) inset). Furthermore, current vs. voltage (I-Vs) were measured at different fields at $T = 0.55$ K, as shown in Fig. 3(d) inset. At $B = 0$, the I-V is linear. For $B > 6$ T, the I-Vs show non-linear behaviors, i.e. the conductance is small at low bias voltages and increases as the voltage increases. Fig. S11 shows the temperature dependence of the I-Vs which demonstrates the melting of the insulating state with increasing temperatures. Our nonlinear transport results further confirm a gap formation for $B > B_C$.

Longitudinal conductivity ($\sigma_{xx} = \frac{\rho_{xx}}{\rho_{xx}^2 + \rho_{xy}^2}$) and Hall conductivity ($\sigma_{xy} = \frac{\rho_{xy}}{\rho_{xx}^2 + \rho_{xy}^2}$) vs. $B$ are shown in Figs. 3(c) and 3(d), respectively. $\sigma_{xx}$ decreases rapidly with decreasing $T$, consistent with the insulating behavior activated above $B_c$. $\sigma_{xy}$ also decreases rapidly for $B > B_c$ and $T < 4$ K, reaching zero at $B_{CNP}$ and remaining near zero up to $B = 72$ T, demonstrating a zero Hall conductivity. The temperature dependence of $B\sigma_{xy}/e$ (Fig. S5(b)) provides insights into the carrier density difference ($n_e - n_h = B\sigma_{xy}/e$) derived from a semiclassical two-band model [17]. For $B > B_{CNP}$, $B\sigma_{xy}/e$ transitions from electron-dominated to charge-neutral transport for $T < 4$ K. Interestingly, $B_{CNP}$, measured using both pulsed (Fig. 3(b) inset) and superconducting magnets (Fig. S9(b)), exhibits a strong temperature dependence well described by a BCS-like equation [17,41].

Previously, the magnetic field-induced metal-insulator transition in ZrTe$_5$ was attributed to charge-density-wave (CDW) formation driven by electron-phonon interactions [27,42] or a disorder-driven carrier-freeze-out [43]. However, to explain our observations we propose a model incorporating an excitonic instability, as detailed below.

In the QL, the effective dimension of the system becomes 1D. It is well known that the electron interaction in 1D is non-perturbative. For instance, the Fermi liquid is replaced by Luttinger liquid for an arbitrarily weak interaction. Luttinger liquid is a critical phase that is susceptible to gap opening. Such gap opening is highly expected when $\mu$ is swept around the Weyl nodes, where Van Hove singularities exist at the top and bottom of the valence/conduction band.

The low energy model [44–47] for HfTe$_5$ is

$$\mathcal{H}(k) = v\left(k_x \sigma_x \tau_x + k_y \sigma_0 \tau_y\right) + \left[M - \xi\left(k_x^2 + k_y^2\right) + \xi_z k_z^2\right]\sigma_0 \tau_z + \mathcal{H}_U \quad [1]$$

where $\sigma$ and $\tau$ act on spin and orbital degrees of freedom and $\mathcal{H}_U$ denotes the Coulomb interaction. For $B$ along the z direction, the orbital contribution in the $\mu = x, y$ plane can be taken into account through a minimal coupling $k_\mu \to -i\hbar\partial_\mu - \frac{e}{c}A_\mu$. The Zeeman energy is $\mathcal{H}_z = -\frac{1}{2}g_1\mu_B B\sigma_z - \frac{1}{2}g_2\mu_B B\sigma_z\tau_z$, where the $g_2$ term accounts for the disparity of the g-factor for the two orbitals. The orbital coupling quantizes the electron motion in the $xy$ plane into LLs, and only the momentum $k_z$ remains a good quantum number. In the QL, the dispersions for the 0$^{th}$ LBs where all electrons reside contain an electron band with spin up and a hole band with spin down. The effective Hamiltonians take a simple form $\mathcal{H}_{\text{eff}} = (\xi_z k_z^2 + m_c)c^\dagger c + (-\xi_z k_z^2 + m_h)h^\dagger h + \mathcal{H}_U$. One salient feature of $\mathcal{H}_{\text{eff}}$ is that the electron (with creation operator $c^\dagger$) and hole branch (with creation operator $h^\dagger$) have the same band mass, and therefore it is expected that the particle-hole susceptibility dominates other instabilities such as the inter-Weyl nodes instabilities. With Coulomb interaction, the system opens a gap by developing the exciton instability associated with the order parameter $\Delta = \langle c_k^\dagger h_k \rangle$. Since $h_k/c_k$ has spin down/up, $\Delta$ has spin quantum one and does not have center of mass momentum. Therefore, the resulting EI is a triplet exciton preserving the translation symmetry. The triplet exciton can coexist with higher fields than the conventional singlet exciton.

The results of detailed model calculations of Eq. 1 are presented in Fig. 4 and the Supplementary Material section VIII. In the QL, as we increase $B$, the 0$^+$ and 0$^-$ LLs move toward each other as a consequence of opposite spin polarization and form a 1D Weyl mode, consistent with the experimental observation of NLMR in S#2. Concomitantly, $\mu$ is lowered from the 0$^-$ LL due to the increased LLs degeneracy and approaches the band edge of the 0$^+$ LL. At low temperatures, the spin-triplet exciton opens a gap $\Delta \neq 0$ in the spectrum. The *T-B* phase diagram for the exciton phase is displayed in Fig. 4d. The triplet excitonic phase shrinks when the Coulomb interaction is suppressed (Fig. S13), i.e. in higher carrier-density systems due to screening. At higher fields, the EI becomes unstable as $\mu$ shifts away from the Weyl node region. However, for a strong Coulomb interaction, the EI phase persists to high fields. Deep in the EI, the electron and hole carriers are compensated, resulting in a vanishing Hall conductivity, which is consistent with the experimental observation of a zero Hall conductivity, as shown in Fig. 3(d) and Fig. S8.

One important manifestation of the exciton phase is the nonlinear electrical transport. At zero temperature and a small bias voltage, the current vanishes because excitons are charge-neutral. When the voltage exceeds a threshold value, excitons are dissolved, and the current jumps to a finite value [48]. At nonzero temperatures, the jump in current is smeared out by thermal fluctuations. The results shown in Fig. 3(d) inset and Fig. S11 align well with the EI picture.

Excitonic states have been reported in systems such as bilayer structures [49–52], monolayer tellurides [53–56], 1T-TiSe$_2$, and Ta$_2$NiSe$_5$ [57,58], often without the need for magnetic fields. In these materials, excitonic order often coexists with charge-density-wave phases, making it challenging to distinguish between the two. This has motivated the search for excitonic materials with valence and conduction bands located at the Brillouin zone center, with a direct gap, avoiding CDW coexistence [59]. Additionally, materials with spin-polarized valence and conduction bands have been predicted to host spin-triplet EIs [36–38], though experimental confirmation remains elusive. HfTe$_5$, with its small Dirac gap centered at the $\Gamma$ point, provides a promising platform to explore spin-triplet EIs in the ultra-QL, as only the spin-polarized $0^-$ and $0^+$ LBs are populated, as illustrated in Fig. 1(a). As $B$ increases, the Fermi wavevector of the electron/hole states overlap at the same momenta, resembling the behavior of a direct-band gap semiconductor. This contrasts with materials like graphite, where carrier pairs exhibit non-zero average momentum at high fields, resembling an indirect-band gap semiconductor [17]. In graphite, the EI phase emerges under high fields [15,17,60–62], with distinct phases identified based on degeneracy lifting in the zero$^{th}$ and first LLs [61].

At the CNP, electron- and hole-like carriers coexist in the LLs, providing favorable conditions for EI formation [63,64]. While a competing CDW phase could form by condensing electron-hole pairs, it would result in continuous Hall conductivity variations with $B$, inconsistent with our observation of vanishing Hall conductivity over a wide field range. Furthermore, the direct-gap nature of HfTe$_5$ disfavors CDW formation [65]. The temperature dependence of $B_{CNP}$ in S#3 (Fig. 3(b) inset and Fig. S9(b)) fits well to a BCS-like model including pair-breaking effects, further supporting the EI scenario.

The spin-triplet EI phase in HfTe$_5$ is analogous to that in the quantum Hall bilayers [66], where excitons form between electrons and holes in LLs. In this case, the layer pseudospin of quantum Hall bilayers is replaced by the physical spin of the excitons, opening possibilities for future studies of phenomena such as Coulomb drag with spin-polarized currents and Josephson-like tunneling effects.

In conclusion, our magneto-transport experiments under high fields on HfTe$_5$, combined with the material-specific modelling, uncover novel emergent quantum phases inherent to its unique topological nature. HfTe$_5$ in the ultra-QL exhibits a simple band structure where $\mu$ intersects only the spin-polarized zero$^{th}$ LBs. With increasing $B$, these bands cross, forming a 1D Weyl phase, evidenced by the NLMR. With further enhanced confinement of the 1D electron liquid by $B$, we observe a metal-to-insulator transition associated with a gap opening (~ 250 μeV). Furthermore, the persistence of a charge-neutral insulating state up to $B$ = 72 T suggests the formation of a correlated phase consistent with a spin-triplet BCS-like EI. Material-specific modeling supports this interpretation, with additional evidence provided by the nonlinear I-V measurements. Collectively, our results provide the magneto-transport evidence for a topological Lifshitz transition and the realization of a spin-triplet EI phase in the ultra-QL of HfTe$_5$, paving the way for exploring exotic correlation-driven phenomena, including spin transport, in 3D topological materials under extreme conditions.

**Acknowledgments**

This research was primarily supported by the Laboratory Directed Research and Development program of Los Alamos National Laboratory under project number 20230014DR. This research was partially supported by the National Science Foundation Materials Research Science and Engineering Center program through the UC Irvine Center for Complex and Active Materials (DMR-2011967). L.A.J. acknowledges the support from NSF-CAREER (DMR 2146567). This work was performed, in part, at the Center for Integrated Nanotechnologies, an Office of Science User Facility operated by the U.S. Department of Energy (DOE) Office of Science. Los Alamos National Laboratory, an affirmative action equal opportunity employer, is managed by Triad National Security, LLC for the U.S. Department of Energy's NNSA, under contract 89233218CNA000001. A portion of this work was performed at the National High Magnetic Field Laboratory, which is supported by National Science Foundation Cooperative Agreement Nos. DMR-1644779, DMR-2128556, the State of Florida and the U.S. Department of Energy. J.L. and L.A.J are grateful to Javier Sanchez-Yamagishi and his group at UCI for the initial measurements at low magnetic fields in his laboratory. We thank Qiyin Lin for assistance with the metal evaporation using the Angstrom e-beam evaporator at the UC Irvine Materials Research Institute (IMRI); and Matthew Law and Geemin Kim for helping us using their thermal metal evaporator. J.L. would like to thank Yi-Xiang Wang at Jiangnan University and Xiang Yuan at East China Normal University for fruitful discussions. We thank Pulsed Field Facility staff Johanna Palmstrom for discussing the data.

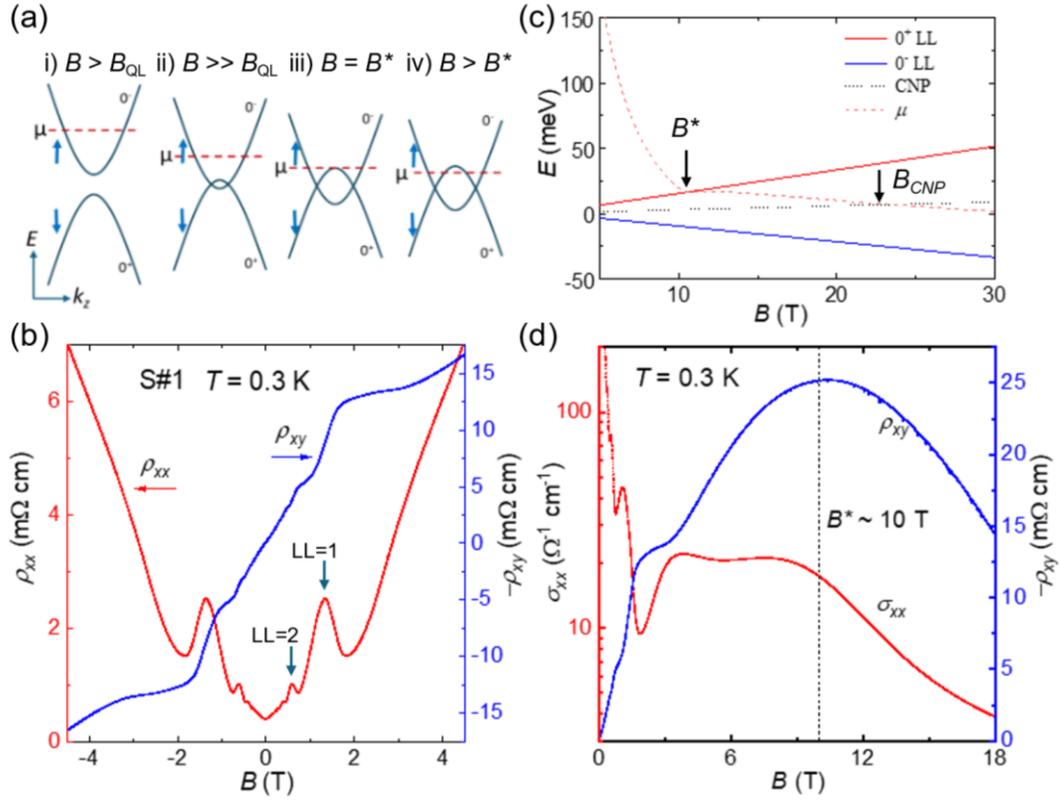

FIG. 1. (a) Schematic of zero[th] LBs and chemical potential ($\mu$) under increasing magnetic field ($B$). (b) $\rho_{xx}$ and $\rho_{xy}$ vs. $B$ measured from S#1 (c) Schematic of the calculated field dependence of zero[th] LBs, $\mu$, and the charge neutrality point (CNP) at $k_z = 0$ for S#1. (d) $\sigma_{xx}$ and $-\rho_{xy}$ vs. $B$. $B^*$ represents the onset of the 1D Weyl mode formation, as shown in (a) iii).

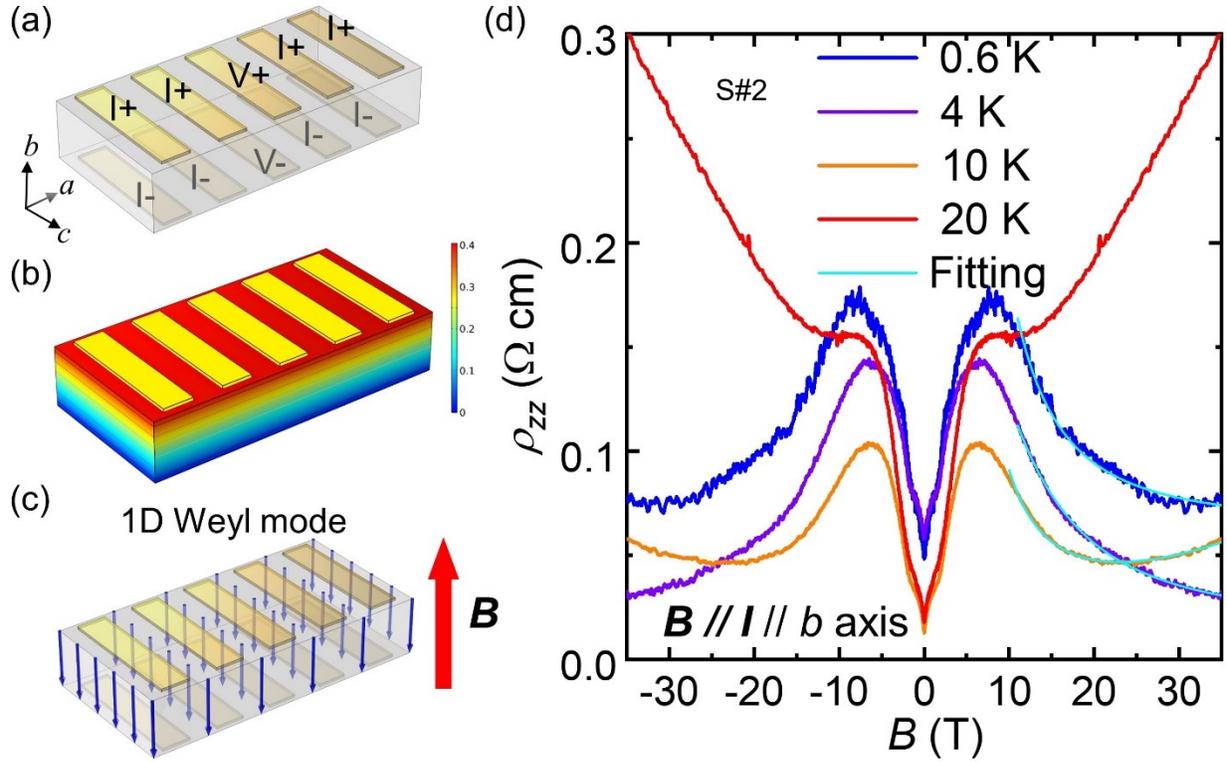

FIG. 2. (a) Schematic of the interlayer resistivity, $\rho_{zz}$, measurement used for S#2. (b) Finite element analysis simulation of the current density in S#2. (c) Schematic of the 1D Weyl mode for $B > B^*$. (d) $\rho_{zz}$ vs. $B$ measured at various temperatures with $B // I // b$. The light blue lines represent the fit for the negative longitudinal magnetoresistance (NLMR) using $\rho_{zz} \propto 1/B^2$. For $T = 10$ K, the fitting also includes a positive magnetoresistance term proportional to $B^2$.

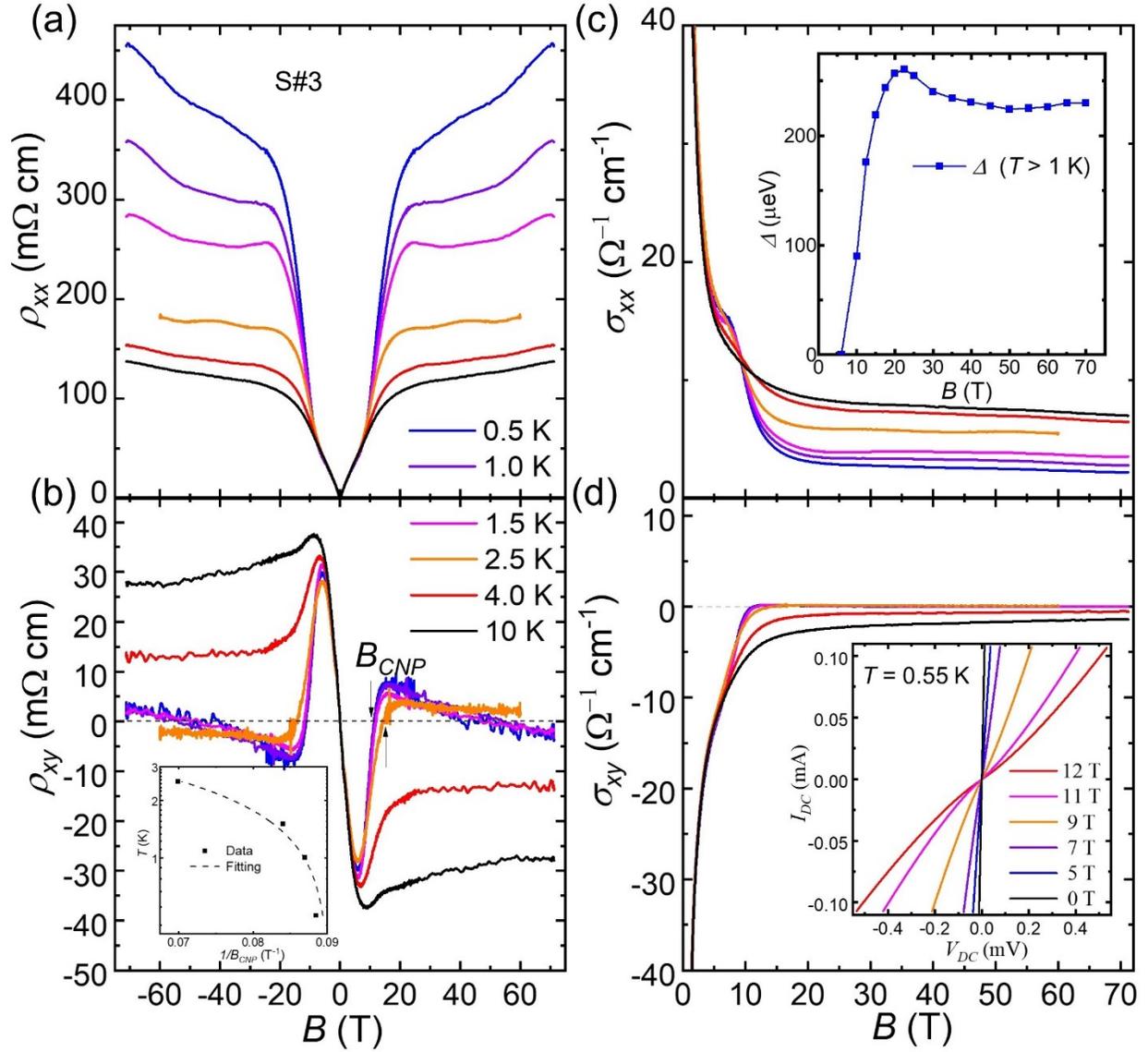

FIG. 3. (a) & (b) $\rho_{xx}$ and $\rho_{xy}$ vs. $B$, at various temperatures as indicated for S#3. Inset: $T$ vs. $1/B_{CNP}$. The fitting is described by Eqs. (1)&(2) in the Supplementary Material. (c) & (d) $\sigma_{xx}$ and $\sigma_{xy}$ vs. $B$ obtained from both $\rho_{xx}$ and $\rho_{xy}$. Inset of (c): activation energy ($\Delta$) vs. $B$, extracted from the temperature dependence of $\rho_{xx}$. Inset of (d): current vs. voltage (I-Vs) measured at $T = 0.55$ K and different representative $B$'s.

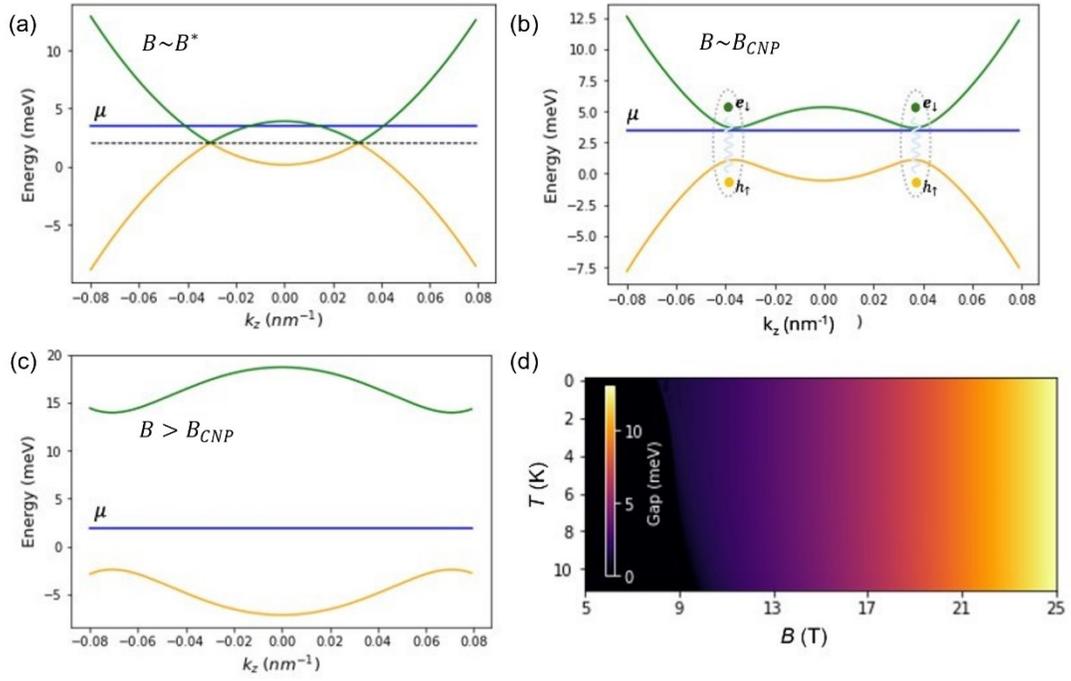

FIG. 4. (a) At $B \sim B^*$, the chemical potential ($\mu$) intersects with the 1D Weyl nodes. (b) and (c) As $B$ increases, electrons and holes form excitons and condense, leading to a gap opening. The system maintains charge neutrality after the exciton condensation. (d) Temperature-magnetic field phase diagram for the exciton insulating phase, where the exciton gap size is shown with the color scale.